\documentclass[letterpaper, 10 pt, conference]{ieeeconf}  

\IEEEoverridecommandlockouts                              
\usepackage{graphicx} 
\usepackage{bm}
\usepackage{cite}

\title{How the nature of web services drives vocabulary creation in social tagging}

\author{Koya Sato$^{1}$, Mizuki Oka$^{2}$, Yasuhiro hashimoto$^{2}$, Takashi Ikegami$^{3}$ and Kazuhiko Kato$^{2}$
\thanks{$^{1}$ University of Tsukuba, PH.D program in Empowerment Informatics, Tsukuba, 305-8573, Japan}%
\thanks{$^{2}$ University of Tsukuba, Faculty of Engineering, Information and Systems, Tsukuba, 305-8573, Japan}%
\thanks{$^{3}$ University of Tokyo, Graduate School of Arts and Sciences, Tokyo, 153-8902, Japan}
}

\begin{document}

\maketitle
\thispagestyle{empty}
\pagestyle{empty}

\begin{abstract}

A social tagging system allows users to add arbitrary strings, called ``tags'', on a shared resource to organize and manage information. The Yule--Simon process, which has shown the ability to capture the population dynamics of social tagging behavior, does not handle the mechanism of new vocabulary creation because it assumes that new vocabulary creation is a Poisson-like random process. In this research, we focus on the mechanism of vocabulary creation from the microscopic perspective and discuss whether it also follows the random process assumed in the Yule--Simon process. To capture the microscopic mechanism of vocabulary creation, we focus on the relationship between the number of tags used in the same entry and the local vocabulary creation rate. We find that the relationship is not the result of a simple random process, and differs between services. Furthermore, these differences depend on whether the user's tagging attitudes are {\it private} or {\it open}. These results provide the potential for a new index to identify the service's intrinsic nature.
\end{abstract}

\section{Introduction}

Online resource sharing services such as {\sf Delicious}\footnote{http://delicious.com/}, {\sf Flickr}\footnote{https://www.flickr.com/}, {\sf Instagram}\footnote{https://www.instagram.com/} and {\sf Twitter} \footnote{https://twitter.com/} have adopted social tagging systems in which users of each service can add an arbitrary string, namely a ``tag'', to a shared resource such as bookmarked URL, text, photo or movie, to organize and manage information~\cite{gupta2010survey,golder2006usage}. While the user's purpose in tagging differs according to the user or service~\cite{strohmaier2012understanding}, a typical use of tagging is to facilitate a user’s future retrieval of information from the vast amounts of continuously increasing resources.

Analysis of tagging behaviors, such as how often new tags are created or how many times they are used, offers insights into the mechanism of the social tagging ecosystem. Indeed, many researches have studied these tagging behaviors~\cite{gupta2010survey,strohmaier2012understanding,trattner2015modeling}. The first attempt to understand tagging behaviors was based on ``Polya's urn model'', proposed by Golder and others~\cite{golder2006usage}. They discovered that social tagging behaviors follow a process known as the ``preferential attachment'' mechanism, according to which the tags that are used more frequently are more likely to be selected. Cattuto and others showed that the relationship between the frequency of tags and their rank, as well as the growth of the vocabulary size, follow power-law forms (i.e., Zipf's law and Heaps' law, respectively)~\cite{zipf1949human,heaps1978information}. These are well-known statistical features observed in the dynamics of social tagging systems. To mathematically model the emergence of these power-law relationships, the Yule--Simon process was introduced~\cite{cattuto2007vocabulary,cattuto2007semiotic}. This discrete stochastic model was originally developed by Yule to explain the power-law population distribution seen in biological species in each genus~\cite{willis1922age,yule1925mathematical}; it was later modified by Simon to apply to other systems, including social phenomena~\cite{1955}. Indeed, many models have been proposed based on ideas of vocabulary growth and preferential attachment~\cite{cattuto2007semiotic,halpin2007complex,dellschaft2008epistemic}.

While the Yule--Simon process nicely describes macroscopic statistical features of tagging behavior, this model ignores some microscopic points of view, namely irregularities or personalities observed in vocabulary creation. The model assumes the vocabulary creation rate to be constant or a function of time; in other words, the Yule--Simon process does not consider the mechanism by which new vocabularies are created. We propose that vocabulary creation does not follow a simple Poisson-like process; rather, the process whereby new vocabularies are created within a similar time period involves certain correlations; this can be semantic, contextual, or related to communication. Based on a similar concept, Tria and others argued for the existence of a correlation among ``novelties'' relating to the concept of the ``adjacent possible'' by Kauffman~\cite{kauffman1996investigations}, which postulates that a novelty appearing in a system becomes a driving force to create another novelty~\cite{tria2014dynamics}. In this paper, we tackle the question from another point of view, namely the correlation that emerges when multiple tags are assigned to a single resource in the presence of a service's intrinsic nature. To accomplish this, we investigate the relationship between the simultaneous use of tags and the local rate of the vocabulary creation in each resource.

This paper consists of four primary sections (in addition to the introduction): We begin with a brief review of the Yule--Simon process to suggest the basis for our discussion. We next explain our experimental methodologies. We then present the results of our empirical analysis and provide an interpretation. In the final section, we discuss our conclusion.

\section{The windowed tag sequence}

\begin{figure*}[htbp]
\centering\includegraphics[width=5.5in]{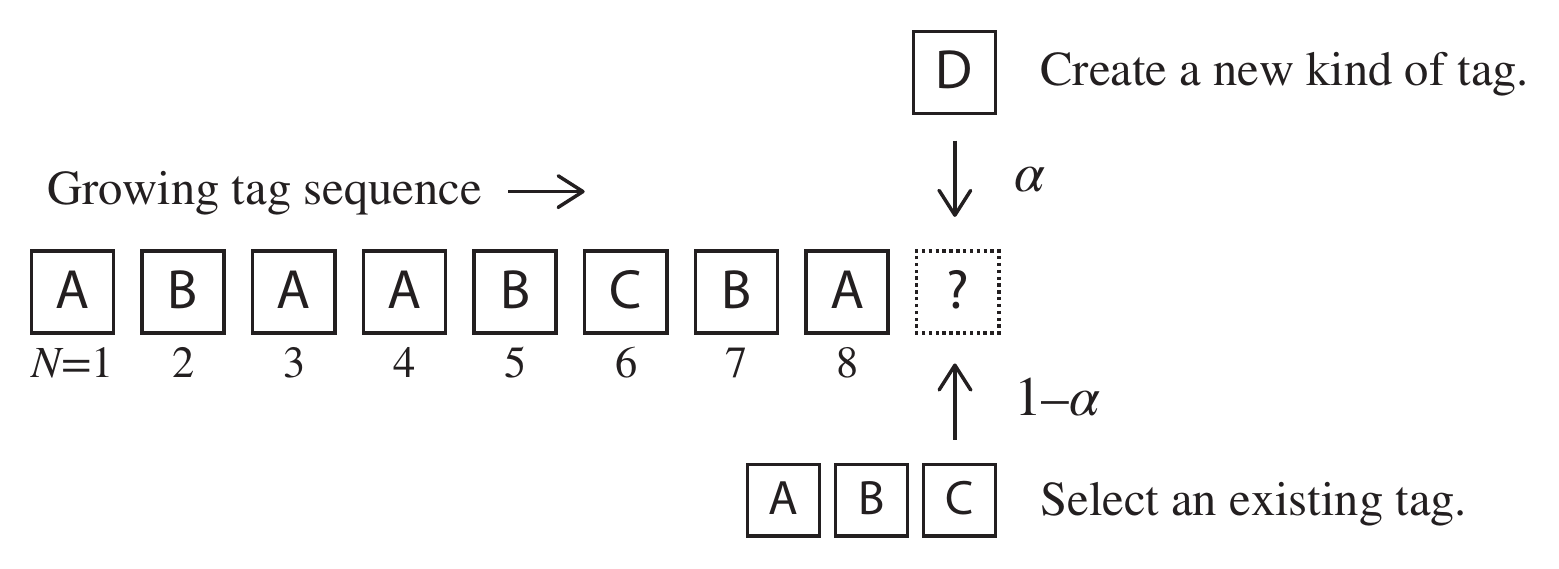}
\caption{Diagram of the Yule--Simon process.}
\label{fig:yule_simon}
\end{figure*}

A schematic diagram of the Yule--Simon process is shown in Fig.~\ref{fig:yule_simon}.
The process evolves in a discrete time step and generates a tag sequence; a tag is added to the sequence in each time step. With a probability $\alpha$, the added tag is a new tag that never appeared in the sequence; with the complementary probability $1-\alpha$, the tag is randomly chosen from among the existing tags in the sequence.

Let us denote $N$ as the total number of tags in the sequence and $k_i$ as the number of tag $i$ appearing in the sequence. Note that $i$ is the index of the distinct tag (vocabulary) and not the index of the individual tag use (annotation). Figure.~\ref{fig:yule_simon} shows the case of N=8, where the tag sequence ({\sf A}, {\sf B}, {\sf A}, {\sf A}, {\sf B}, {\sf C}, {\sf B}, {\sf A}) is realized. The probability of the existing tag $i$ being chosen, $P(i)$, in the $(N+1)$th step is given by the following:
\begin{equation}
\label{eq_1}
P(i)=(1-\alpha)\frac{k_i}{N}.
\end{equation}
If the existing tag is chosen uniformly at random from the sequence in step $N=9$, the next step in time, the resulting probabilities of each vocabulary to be chosen are $4/8$, $3/8$ and $1/8$ {\sf A}, {\sf B} and {\sf C}, respectively. That is, there is a positive feedback known as preferential attachment working in tag selection.

Considering the multiple tag assignment to a single resource in actual social tagging behavior, we introduce an idea of a ``window''. Each entry that people post to the service is defined as $e_i = \{u_i, r_i, {\bm t}_i\}$, where the index $i$ is given in ascending order of time throughout all entries, $u_i$ is the user who posted the $i$-th entry, $r_i$ is the target resource, and ${\bm t}_i$ is the set of tags assigned to $r_i$ by $u_i$. We call the set of tags in the entry the window and denote the window size of the $i$-th entry as $w_i = |{\bm t}_i|$. In the ordinary Yule--Simon process, each tag is added to the sequence, one by one, following the probabilistic trial defined above. However, each entry is now generated in every time step, and the sequence of tags grows according to the window size. Herein, we see two kinds of tag series: One is microscopically generated in a window by a single person; the other is macroscopically realized by successive windows. In other words, the macroscopic one is successive entries by multiple folks. In Eq.(\ref{eq_1}), $\alpha$ is given as a single parameter or a time-dependent, gradually decreasing function in the ordinary Yule--Simon process. However, $\alpha$ is now supposed to be different from window to window and person to person in actual social tagging behavior. Our principal interest is in revealing the relationship between this window-specific $\alpha$ and the statistics of the window.

\section{Measuring users' tagging motivation}
It is important to consider the tagging motivation of users. We assume that tagging motivation governs, to a certain extent, the window statistics and corresponding $\alpha$. Let us recall that tags are used for different purposes in different services. They might be used, for example, to share resources with others or for more personal use~\cite{trant2009studying,hammond2005social,heckner2009personal}. Consequently, it is supposed that the statistics of the window and $\alpha$ also differ according to the specific service. Heckner and others argued that the number of tags assigned to a single resource, that is, the window size here, is affected by the tagging motivation~\cite{heckner2008tree}. If a user employs tags with the purpose of sharing resources with others, the window size becomes large; this is called ``over-tagging''~\cite{heckner2008tree}. It is natural to assume that as the number of tags assigned to a resource increases, the ease with which people can reach that resource by tag query increases as well. We call tagging behavior based on this purpose an {\itshape open} attitude. In contrast, a {\itshape private} attitude relates to more personal use. Thus, we measure window size to reveal the motivation behind each service and each entry. 

While the above interpretation regarding the size of the window is reasonable, this value does not address the difference in user tagging preferences when users choose tags. Treating this aspect of tag selection numerically, Stromaier and others identified two typical personalities in social tagging: ``describer'' and ``categorizer''~\cite{strohmaier2012understanding,korner2010categorizers}. We associate the behavior of describers and categorizers with {\it open} and {\it private} attitudes in tagging, respectively, for the following reason: A categorizer is defined as a user who uses the same vocabulary repeatedly over different resources for personal information retrieval. In contrast, a describer is defined as a user whose tag assignment is biased, as compared to the tag assignment of categorizers, in order to precisely describe individual resources. Stromaier and others defined two indexes, $M_0$ and $M_1$, to distinguish each user as a describer or a categorizer. $M_0$ is defined as follows:
\begin{equation}
M_0=\frac{\left| \{t:|{\bm R}(t)| \leq n \}\right|}{|{\bm T}|},\quad n = \left\lceil \frac{|{\bm R}(t_{\max})|}{100} \right\rceil,
\end{equation}
where $|{\bm T}|$ is the vocabulary size, $|{\bm R}(t)|$ is the number of resources to which tag $t$ is assigned, and $|{\bm R}(t_{\max})|$ is the number of resources to which the most used tag is assigned. $M_0$ ranges $[0:1]$ and takes a large value when the user employs various tags but uses each tag just a few times. Such users are considered describers. $M_1$ is defined as follows:
\begin{equation}
M_1 = \frac{H({\bm R}|{\bm T})-H_{\mbox{opt}}({\bm R}|{\bm T})}{H_{\mbox{opt}}({\bm R}|{\bm T})},
\end{equation}
where $\bm T$ is a set of tags, $\bm R$ is a set of resources and $H({\bm R}|{\bm T})$ is the conditional entropy, which captures the bias of the number of resources per tag. $H_{\mbox{opt}}({\bm R}|{\bm T})$ is an optimal value of $H({\bm R}|{\bm T})$ defined for the ideal categorizer, and it works as a normalization factor. $M_1$ ranges $[0:\infty]$ but is less than 1 in most cases. A small value of $M_1$ suggests that the user can be considered a categorizer. Finally, the averaged value of these two indexes,
\begin{equation}
\label{eq_m}
M = \frac{M_0+M_1}{2},
\end{equation}
is actually used to identify the user as either a describer or a categorizer. When the value is close to 0, the user is identified as a categorizer; otherwise, the user is identified as a describer. According to Stromaier and others, the tags used by describers tend to be more consensual, that is, common vocabularies among all users, compared with those used by categorizers~\cite{strohmaier2012understanding}. This observation also supports our assumption that describers use tags in an {\itshape open} manner, while categorizers do so in a {\itshape private} manner. We use the two indexes described here to numerically identify {\itshape open} and {\itshape private} tagging motivation.

\section{Empirical analysis}
We analyze the actual social tagging data based on the ideas mentioned above. First, we investigate the existence of a correlation between the vocabulary creation rate, $\alpha$, and the window size, $w$. Next, we identify the user's tagging motivation by measuring the window size distribution and Eq.~\ref{eq_m}, and discuss the relationship between such motivation and the $\alpha$-$w$ correlation.

\subsection{Datasets}

\begin{table*}[!ht]
\begin{center}
\caption{The four datasets in numbers.}
  \begin{tabular}{l|r|r|r|r|r} \hline \hline
    Service & Users & Vocabularies & Annotations & Entries & Range\\
    \hline
{\sf Delicious} & 532,924 & 2,481,108 & 140,126,555 & 47,257,452 & Jan. 2003 - Dec. 2006\\
{\sf Flickr} & 319,686 & 1,607,879 & 112,900,000 & 28,153,045 & Jan. 2004 - Dec. 2005\\
{\sf Instagram} &2,110 &271,490 &8,201,542&1,047,774 & Oct. 2010 - Feb. 2014\\
{\sf RoomClip} &32,852 &194,881 &3,141,524& 692,459 & Apr. 2012 - May. 2015\\
\hline \hline
  \end{tabular}
  \end{center}
  \label{table:data}
\end{table*}

We examined four datasets obtained from four web services— {\sf Delicious}, {\sf Flickr}, {\sf Instagram} and {\sf RoomClip} \footnote{https://roomclip.jp/}-that use a social tagging system. The first service ({\sf Delicious}) is used to share web bookmarks; the other three services are used to share photos. Users of {\sf Delicious} mainly use the service for bookmarking and preserving data, while {\sf Instagram} is typically used to share a user’s own resources with other users\cite{Difference,heckner2009personal}. {\sf Flickr} lies midway between {\sf Delicious} and {\sf Instagram}\cite{Difference, heckner2009personal}. Because {\sf RoomClip} has not yet attracted the attention of researchers, the predominant usage pattern for this service has not been clearly identified. However, we regard it as similar in nature to {\sf Instagram}, rather than {\sf Flickr}, since, as is true for {\sf Instagram}, an uploaded photo in {\sf RoomClip} is automatically accessible to all users. Also, unlike {\sf Flickr}, {\sf RoomClip} has no storage limitation for uploaded photos. These characteristics should affect a user's tagging behavior. In the following section, we numerically characterize the dominant tagging behavior for each service.

The four data sets were assembled in different ways. The {\sf Delicious} and {\sf Flickr} data were gathered by G\"{o}rlitz and others by massively scraping the two services, which means the data should be fairly comprehensive~\cite{gorlitz2008pints}. {\sf Instagram} data were gathered by Ferrara and others from over 2,000 specific users who were randomly selected among users who participated in specific events~\cite{ferrara2014online}. {\sf RoomClip} data were directly provided by Tunnel Inc.\footnote{http://www.tunn-el.com/}, which runs the service, so that the data collected here covers the tagging behavior of all its users~\footnote{In {\sf RoomClip} data, there is a set of nine specific tags defined by the company. 
The user is required to choose one of these tags before adding new tags.
We removed the nine authoritative tags from our analysis.}. Moreover, this data is anonymized. In {\sf Flickr}, {\sf Instagram} and {\sf RoomClip}, each entry is uniquely associated with the resource. In contrast, {\sf Delicious} is based on a {\itshape broad tagging system}~\cite{broad_narrow}, where multiple users can tag to the same resource; therefore, each entry is identified by a user ID and resource ID pair. However, this difference has no significance in our analysis. Table 1 shows the relevant numbers for these four datasets.

\subsection{The correlation between $\alpha$ and $w$}
\begin{figure*}[!h]
\centering\includegraphics[width=6.0in]{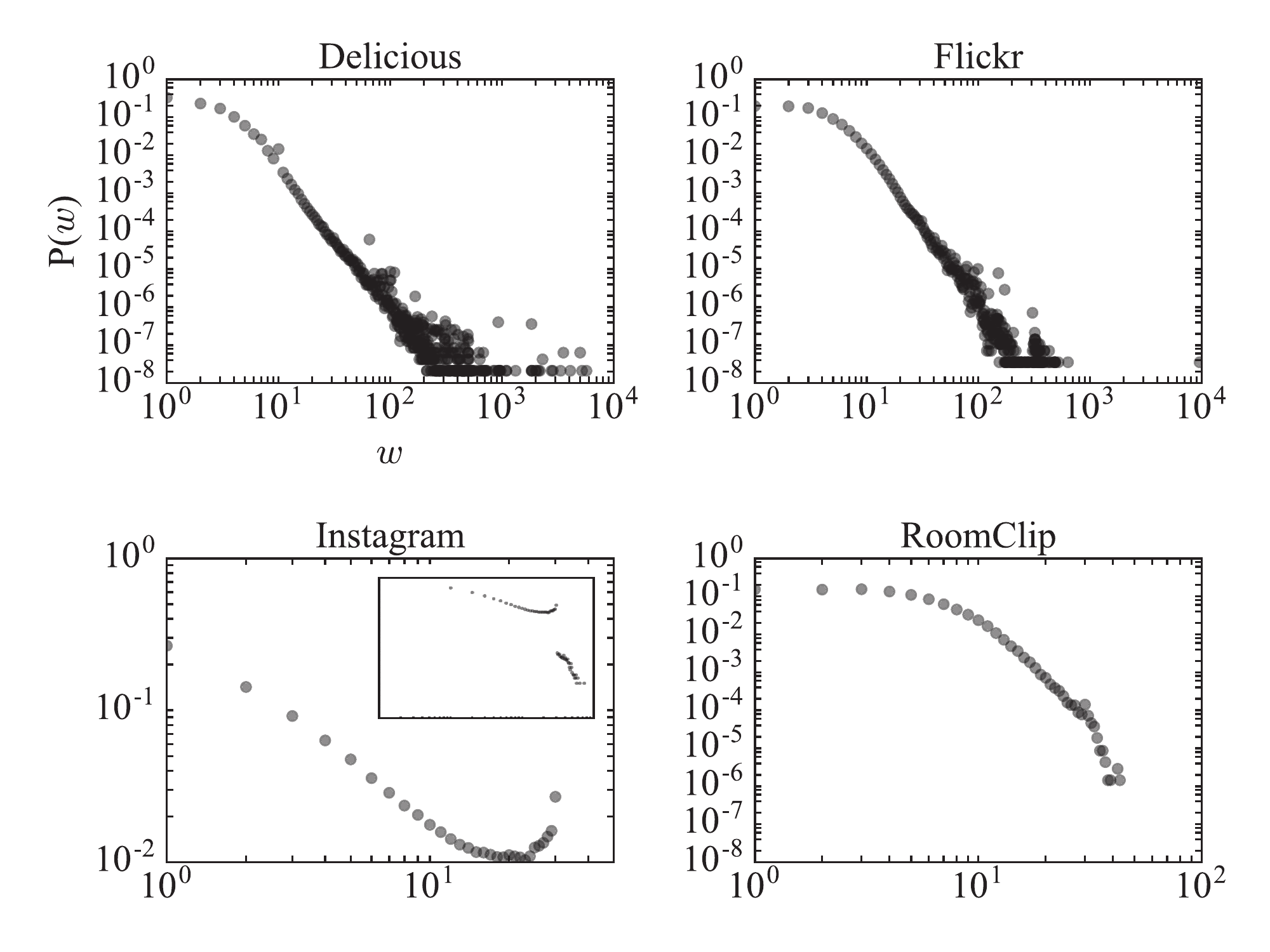}
\caption{The probability distribution of $w$ for each service. In the case of {\sf Instagram}, we focus on the range of $w$ between 1 and 30. The insert for {\sf Instagram} shows the entire $w$ distribution.}
\label{fig:w_dist}
\end{figure*}
Figure~\ref{fig:w_dist} shows the $w$ distribution for the four datasets. {\sf Delicious} and {\sf Flickr} show a power-law distribution, meaning the range of $w$ is wide and the volume of entries drastically decreases as $w$ increases. While the range of $w$ for the {\sf Instagram} and {\sf RoomClip} data is narrower than the range for {\sf Delicious} and {\sf Flickr}, {\sf Instagram} and {\sf RoomClip} also show a power law-like distribution, which, again, means that the volume of entries with a large $w$ drastically decreases with an increase of $w$. In the case of {\sf Instagram}, we focus on the range of $w$ between 1 and 30 since there is a gap at $w=30$ caused by  an inherent {\sf Instagram} penalty whereby users are temporarily prevented from commenting on content on which tags are put more than $30$ times. The insert shows the entire $w$ distribution for {\sf Instagram}.

\begin{figure*}[!h]
\centering\includegraphics[width=6.0in]{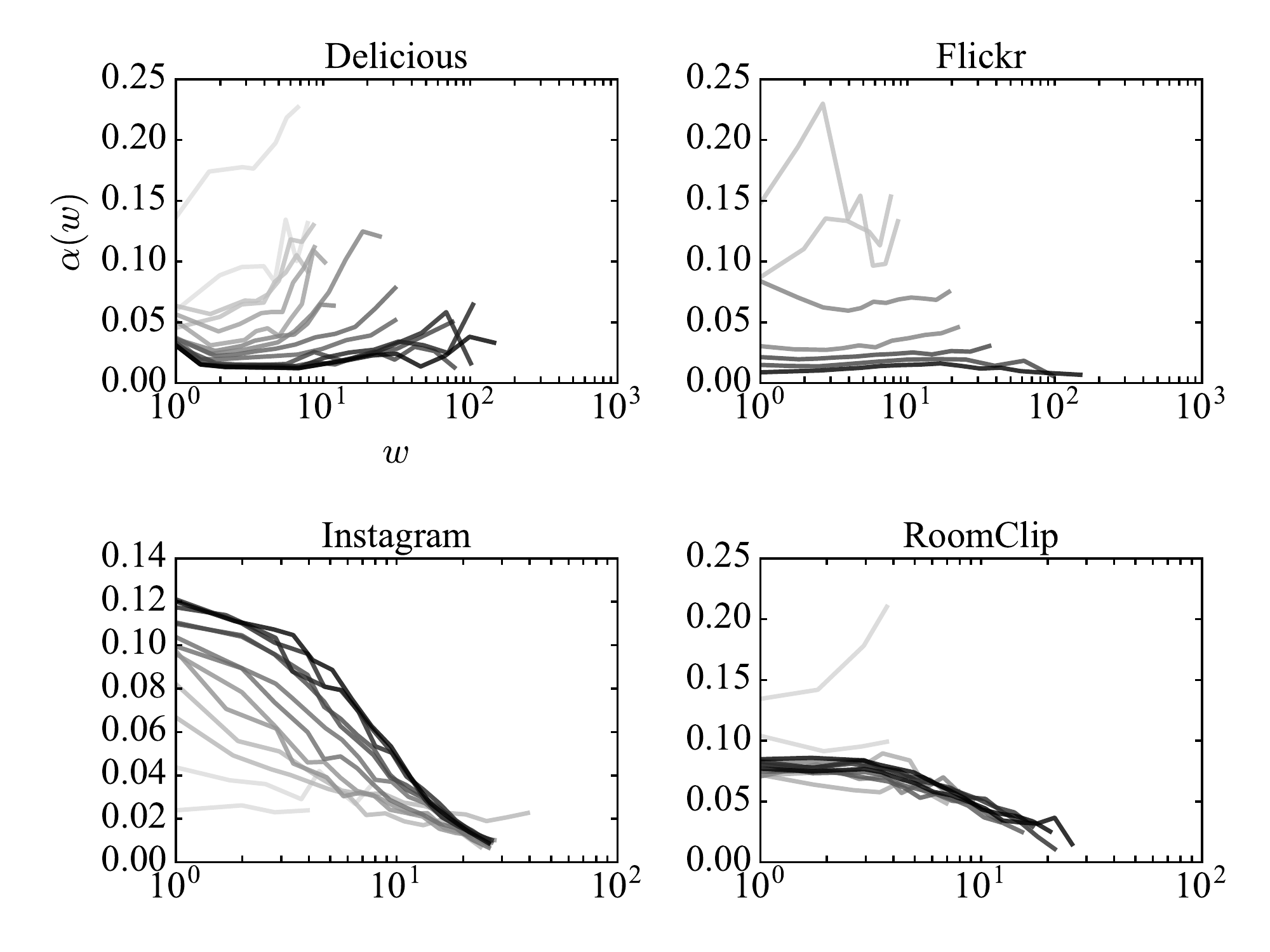}
\caption{The correlation between $\alpha$ and $w$. The horizontal axis is the window size; the vertical axis is the mean vocabulary creation rate for each bin of a window. The color gradation shows the index for the period.}
\label{fig:correlation}
\end{figure*}

Figure~\ref{fig:correlation} shows the correlation between $\alpha$ and $w$ for the four datasets. The sequence of entries is divided into three-month periods for each service, since it is assumed that the $\alpha-w$ correlation gradually becomes stable with the increase of entries and vocabulary. This means that in the initial stages of a service there is not sufficient vocabulary to properly consider; in this stage, almost all the vocabulary that appears is regarded as new vocabulary. We express this notion with a color gradation in the figure. The color gradually becomes black as time elapses over successive periods. Accordingly, we can now denote $\alpha$ and $w$ for the $i$-th entry in each period, by using $\alpha_{i}$ and $w_{i}$, respectively. Note that new vocabulary means that the vocabulary appears for first time, including previous periods. Also, the window size is split into the same-size 20 log-scale bins for each period, $b(w_{i})=\lfloor 20\frac{\ln(w_{i})}{\ln(\mbox{max}({w}))}\rfloor$, so that $w$ has the power law-like distribution. We average the $\alpha$ for each bin as follows:
\begin{equation}
\alpha(w)=\frac{\sum_{j\in {\bm e}(w)}\alpha_{j}}{|{\bm e}(w)|},\quad {\bm e}(w)=\{i\mid b(w_{i})=b(w)\},
\end{equation}
where ${\bm e}(b)$ represents the entries whose window size falls in the same bin as $w$. We ignore any bin where the number of entries does not exceed 100 because the average $\alpha$ of such bins is unstable and noisy. From the figure, we can see that the relationship between window size and vocabulary creation rate exhibits two different patterns of growth in a period. The {\sf Delicious} and {\sf Flickr} data show a weak positive correlation and no-correlation, respectively, while the {\sf Instagram} and {\sf RoomClip} data show a negative correlation. The existence of correlation suggests that there is another tag creation mechanism at work that is not considered in the Yule--Simon process.

The weak positive and no correlation shown in the {\sf Delicious} and {\sf Flickr} data indicate either that an increase in $w$ is caused by vocabulary creation, or, alternatively, that an increase in $w$ causes vocabulary creation. Intuitively, the first of these two possibilities —that the increase in $w$ is caused by vocabulary creation-seems to make more sense. This would be the result of users acting with a {\itshape private} attitude, where the aim is to categorize resources with high accuracy for private use. In contrast, the negative correlations shown in the {\sf Instagram} and {\sf RoomClip} data indicate that an increase in $w$ is caused by the selection of previously used tags or that increasing $w$ does not lead to the vocabulary creation. This would be the result of {\itshape open} attitude behavior by users, where the aim is to share resources with others; in such cases, it would be nonsensical to use a brand new vocabulary term for the purpose of being searched by others. To understand why the relationships are different for each of the services, we can look at the user's tagging motivation numerically from the aspect of {\itshape open} and {\itshape private} tagging attitudes, as described next.

\subsection{The user's motivation for tagging}
\begin{table}[!ht]
  \label{tab:windowed}
\begin{center}
\caption{The median and mean of $w$ in each service}
  \begin{tabular}{l|r|r} \hline \hline
    Service & Median of $w$ & Mean of $w$\\
    \hline
{\sf Delicious} & 2 & 2.96 \\
{\sf Flickr} & 3 & 4.01 \\
{\sf Instagram} & 3 & 7.82 \\
{\sf RoomClip} & 4 & 4.53 \\
\hline \hline
  \end{tabular}
  \end{center}
\end{table}
Table 2 shows the mean and median of $w$ for each service. Following Heckner's study (mentioned above), we consider that a service showing a large value of $w$ means that the dominant users of the service exhibit {\itshape open} attitude behavior, while a service showing a small value of $w$ means that users exhibit {\itshape private} attitude behavior. Based on our numbers, then, we would expect that users of {\sf Delicious} and {\sf Flickr} generally have a {\itshape private} attitude, as compared with users of {\sf Instagram} and {\sf RoomClip}; on the other hand, we would expect {\sf Instagram} and {\sf RoomClip} users to have an {\itshape open} attitude, as compared with {\sf Delicious} and {\sf Flickr} users.

\begin{figure*}[!h]
\centering\includegraphics[width=6.0in]{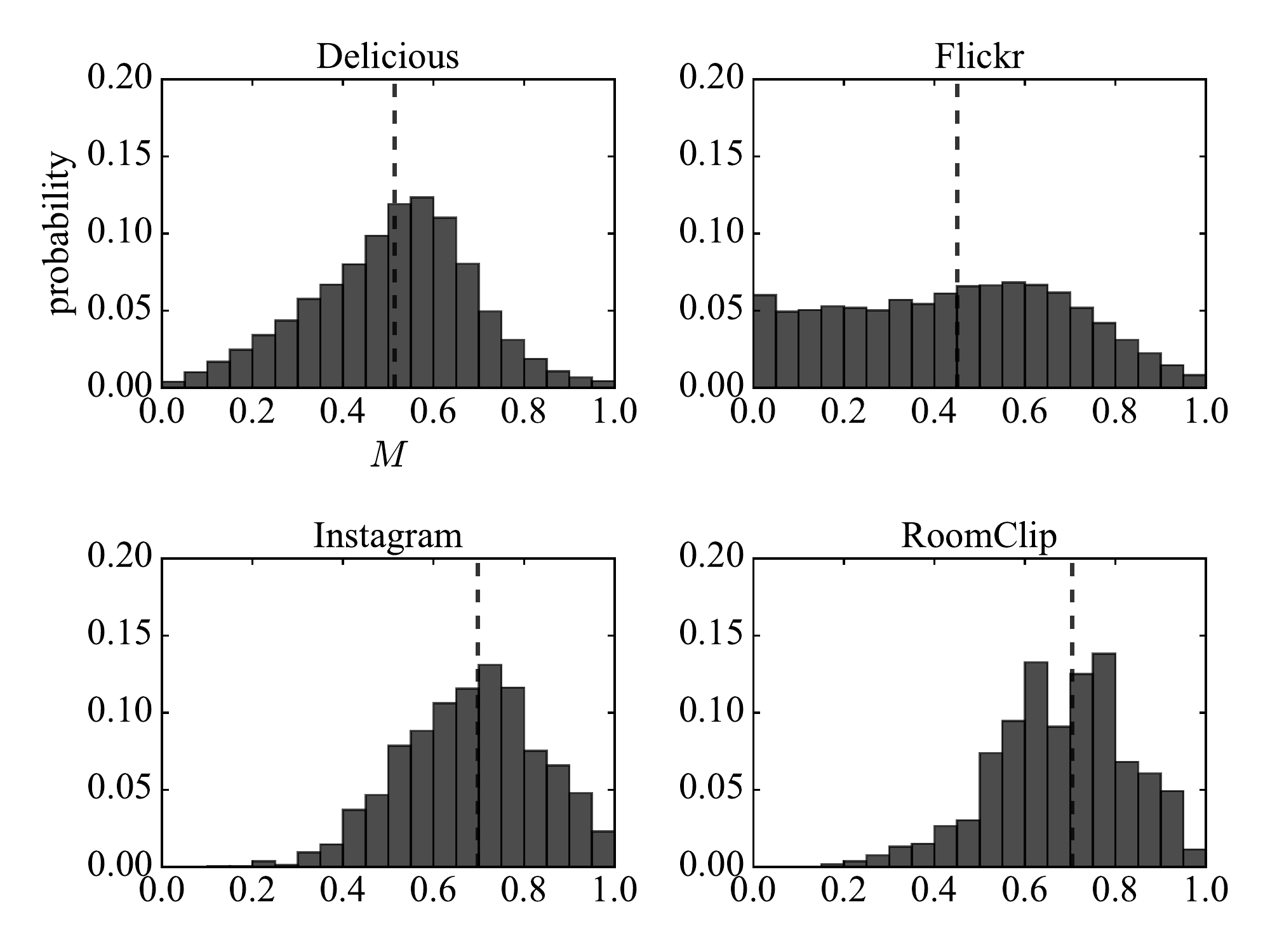}
\caption{The distribution of the index M for each service. Users with $M=0$ are extreme categorizers; users with large $M$ close to 1 are extreme describers.}
\label{fig:desc_categ}
\end{figure*}
We next identify user tagging motivation by calculating the index M defined in Eq.(\ref{eq_m}) for all users. We extracted effective users who posted more than 200 entries and whose $H_{\mbox{opt}}({\bm R}|{\bm T})$ value was not 0, since such users are familiar with the service and, at the same time, the value of $M$ is stable. Figure~\ref{fig:desc_categ} shows the resulting $M$ distribution for each service. The horizontal axis shows the value of $M$, divided into 20 bins in order to construct a histogram. The vertical axis shows the fraction of the number of users in each bin. As described earlier, a user with a small value of $M$ is a ``categorizer''; a user with a large value of $M$ is a ``describer''. The vertical dashed line is the mean value of M. We can characterize the motivation of users by looking at the shape of the distribution. A distribution that is biased to the left indicates that the service is dominated by categorizers; a distribution that is biased to the right indicates that the service is dominated by describers. In Fig~\ref{fig:desc_categ}, Instagram and RoomClip appear to be dominated by describers, as compared to {\sf Delicious} and {\sf Flickr}. That is, user attitudes related to tagging in {\sf Instagram} and {\sf RoomClip} are more open than those in Delicious and {\sf Flickr}. In the case of {\sf Flickr}, the shape of the distribution is relatively flat, which means that the tagging motivations in {\sf Flickr} are broadly distributed, as compared with the other three services. At the same time, we see a volume of strong describers in {\sf Flickr}.

From the above two analyses, we conclude that {\sf Delicious} and {\sf Flickr} are dominated by users whose tagging manner is {\itshape private}. In contrast, {\sf Instagram} and {\sf RoomClip} are dominated by users with an {\itshape open} manner. These results are consistent with our original assessment regarding the main usage of the four services.

\section{Conclusion}
In this paper, we describe the correlation between the number of tags and the vocabulary creation rate in a single resource. This $w-\alpha$ correlation exhibits a different pattern in different services depending on the nature of the service. To explain why these differences arise, we focused on the tagging motivation of users and clarified the relationship between the $w-\alpha$ correlation and tagging motivation resulting in a consistent tagging attitude, namely, the {\itshape open} and {\itshape private} manners in tagging. The weak positive and no $w-\alpha$ correlation observed in {\sf Delicious} and {\sf Flickr} were associated with the {\itshape private} attitude; in contrast, the negative $w-\alpha$ correlations observed in {\sf Instagram} and {\sf RoomClip} were associated with the {\itshape open} attitude. We regard the {\itshape open} and {\itshape private} attitudes as communication-oriented and personal-oriented behavior, respectively. The nature of the associated services seems to be extremely supportive of our interpretation.

To summarize, we argued two points: 1) The vocabulary creation rate in a single resource has a correlation with the number of tags assigned to the resource, and the correlation pattern explains what the nature of the service is; and 2) Conversely, the implication of the vocabulary creation in each service differs according to the intrinsic nature of the service, and considering the number of tags in the resource gives a novel perspective in designing new indexes to identify the service's intrinsic nature.

\section*{ACKNOWLEDGMENT}

This work was supported by the Japan Society for the Promotion of Science(https://www.jsps.go.jp/english/) KAKENHI Grant Number 16K00418, and partially by KAKENHI Grant Number 15K00420. 


\bibliographystyle{abbrv}
\bibliography{main}

\end{document}